\newif\ifcomments\commentsfalse
\newif\iflong \longtrue

\documentclass[runningheads]{llncs}

\usepackage{color}
\usepackage{url}
\usepackage{cite}

\usepackage{color}
\definecolor{dkblue}{rgb}{0,0.1,0.5}
\definecolor{dkgreen}{rgb}{0,0.4,0}
\definecolor{dkred}{rgb}{0.6,0,0}
\definecolor{linkColor}{rgb}{0,0,0.5}
\definecolor{darkblue}{rgb}{0.0,0.0,0.3}

\usepackage{graphicx}
\usepackage{wrapfig}
\usepackage{listings}
\usepackage{xspace}

\newcommand\maybecolor[1]{\color{#1}}
\newcommand\aseem[1]{\ifcomments{\maybecolor{magenta}{\small{[Aseem: #1]}}}\fi}

\let\ls\lstinline

\newcommand{\kw}[1]{\mbox{\normalfont\lstinline!#1!}}
\def\lstcode#1#2#3{\lstinputlisting[linerange=#2-#3]{Code/#1}}
\def\lstfrag#1/#2.{\lstcode{#1}{#2Begin}{#2End}}

\newcommand\citep[1]{\cite{#1}}
\newcommand\citet[1]{\cite{#1}}
\newcommand\ourlang{\textsc{Wys$^\star$}\xspace}
\newcommand\fstar{F$^\star$\xspace}

\newcommand\eat[1]{}
\newcommand\mc{MPC\xspace}

\newcommand{\Paragraph}[1]{\paragraph*{#1.}}

\usepackage{amsmath}
\usepackage{mathpartir}
\usepackage{multirow}
\usepackage{amssymb}

\begin{document}
\title{\ourlang: A DSL for Verified \\Secure Multi-party Computations}

\iflong
\author{Aseem Rastogi\inst{1} \and
Nikhil Swamy\inst{1} \and
Michael Hicks\inst{2}}

\authorrunning{Rastogi et al.}
\institute{Microsoft Research\\
\email{\{aseemr,nswamy\}@microsoft.com} \and
University of Maryland\\
\email{mwh@cs.umd.edu}}
\else
\author{}
\institute{}
\fi




\maketitle

\begin{abstract}
Secure multi-party computation (\mc) enables a set of mutually
distrusting parties to cooperatively compute, using a cryptographic
protocol, a function over their private data. 
%
This paper presents \ourlang, a new domain-specific language (DSL)
for 
writing \emph{mixed-mode} \mc{s}. \ourlang is an embedded DSL hosted in \fstar, a
verification-oriented, effectful programming language.
\ourlang source programs are
essentially \fstar{} programs written in a custom \mc effect, meaning
that the programmers can use \fstar{}'s logic to verify the correctness
and security properties of their programs. To reason about the
distributed runtime semantics of these programs, we formalize a deep
embedding of \ourlang, also in \fstar{}. We mechanize the necessary
metatheory to prove that the properties verified for the \ourlang
source programs carry over to the distributed, multi-party
semantics. Finally, we use \fstar{}'s extraction to
extract an interpreter that we have proved matches this semantics,
yielding a partially verified implementation. \ourlang is the first
DSL to enable formal verification of \mc programs.
We have implemented several \mc protocols in \ourlang, including
private set intersection, joint median, and an \mc-based card dealing
application, and have verified their correctness and security.
\end{abstract}

\section{Introduction}
\label{sec:intro}

Secure multi-party computation (\mc) enables two
or more parties to compute a function $f$ over their private inputs
$x_i$ so that parties don't see each others' inputs, but
rather only see the output $f(x_1,...,x_n)$. Using a trusted
third party to compute $f$ would achieve this goal, but in fact we can
achieve it using one of a variety of cryptographic protocols carried
out only among the
participants~\cite{shamir1980mental, Yao, GMW, BMR}.
One example use of \mc is private set intersection (PSI): the
$x_i$ could be individuals' personal interests, and the function $f$
computes their intersection, revealing which interests the group has
in common, but not any interests that they don't. \mc has also been used for 
auctions~\cite{Bogetoft2009}, detecting tax fraud~\cite{taxfraudsmc},
managing supply chains~\cite{supplychainsmc}, privacy
preserving statistical analysis~\cite{UT:Kamm15}, and more recently
for machine learning tasks~\cite{ezpc,Buscher:2018:HCH:3243734.3243786,gazelle,secureml,Liu:2017:ONN:3133956.3134056}.

Typically, cryptographic protocols expect $f$ to be specified as a
boolean or arithmetic circuit. Programming directly with circuits and
cryptography is painful,
so starting with the Fairplay project~\cite{fairplay} many researchers
have designed higher-level domain-specific languages (DSLs) in which
to program
\mc{s}~\cite{Huang11,viff,Malka2011,fairplaymp,Holzer12,Nielsen07,Nielsen09,sharemind,Schropfer2011,wysteria,Liu2014,Laud:2015:DLL:2810103.2813664,Crockett:2018:ALC:3243734.3243828,Araki:2018:GSC:3243734.3243854,Buscher:2018:HCH:3243734.3243786,frigate}.
These DSLs compile source code to
circuits which are then given to the underlying protocol. While doing this
undoubtedly makes it easier to program \mc{s}, these languages still
have several drawbacks regarding both security and usability.

This paper presents \ourlang, a new \mc DSL that addresses several
problems in prior DSLs. Unlike most previous \mc DSLs, \ourlang is not a
standalone language, but is rather an embedded DSL hosted in
\fstar~\cite{fstar2016}, a full-featured, verification-oriented, effectful
programming language. \ourlang has the following two distinguishing
elements: 

\Paragraph{1. A program logic for \mc} (\S\ref{sec:overview} and \S\ref{sec:formal}.)
In their most general form, \mc applications are \emph{mixed-mode}:
they consist of parties
performing (potentially different)
local, in-clear computations (e.g. I/O, preprocessing
inputs) interleaved with joint, secure computations. \ourlang is the
first MPC DSL to provide a program logic to formally reason about the
\emph{correctness and security} of such applications,
e.g., to prove that the outputs will not reveal too much information
about a party's inputs~\cite{mardziel13belief}.\footnote{Our attacker
  model is the ``honest-but-curious'' model where the attackers are
  the participants themselves,
  who play their
  roles in the protocol faithfully, but are motivated to deduce as much as they
  can about the other participants' secrets by observing the
  protocol. \S\ref{sec:wysts} makes the security model of \ourlang more precise.}

To avoid reasoning about separate programs for each
  party, \ourlang builds on the basic programming model of the
  Wysteria \mc DSL~\cite{wysteria} that allows applications to be
  written as a single specification. \ourlang presents a
\emph{shallow embedding} of the Wysteria programming model in \fstar.
When writing \ourlang source programs, programmers essentially write
\fstar programs in a new \ls{Wys} effect,
against a library of \mc combinators. The pre- and
postcondition specifications on the combinators encode a program logic
for MPC. The logic provides
\emph{observable traces}---a novel addition to the Wysteria
semantics---which programmers can use to 
specify security properties such as delimited release~\cite{sm03delimited}.
Since \ourlang programs are \fstar programs, \fstar computes
verification conditions (VCs) for
them which are discharged using Z3~\cite{z3} as usual.

We prove the soundness of the program logic---that the properties proven about
the \ourlang source programs carry over when these programs are run by
multiple parties in a distributed manner---also in \fstar. The proof connects
the pre- and postconditions of the \ourlang combinators to their distributed
semantics in two steps.
First, we implement the combinators in \fstar, proving the
validity of their pre- and postconditions of against their implementation.
Next, we reason about this implementation and the distributed runtime
semantics through a deepembedding of \ourlang in \fstar.
Essentially, we deep-embed the \ourlang combinator abstract syntax trees (ASTs)
as an \fstar
datatype and formalize two operational semantics for them:
a conceptual single-threaded semantics that models their \fstar implementation,
and the actual distributed
semantics that models the multi-party runs of the
programs. We prove, in \fstar, that the single-threaded
semantics is sound with respect to the distributed semantics (\S\ref{sec:formal}).
While we use \fstar, the program logic is general and
it should be possible to embed it in other verification frameworks
(e.g., in Coq, in the style of Hoare Type Theory~\cite{ynot-icfp08}).

\paragraph*{2. A full-featured, partially verified implementation}
(\S\ref{sec:formal}.) 
\ourlang's implementation is, in part, formally verified. The hope is
that formal verification will reduce the occurrence of security
threatening bugs, as it has in prior
work~\cite{csmith,Leroy2009Formal,BhargavanFKPS13,polarssl,export:122884}.

We define an interpreter in \fstar that operates over the \ourlang
ASTs produced by a custom \fstar
extraction for the \ls{Wys} effect.
While the local computations are executed locally by the interpreter,
the interpreter compiles secure-computation ASTs
to circuits, on the fly, and executes them using the Goldreich, Micali and
Wigderson (GMW) multi-party computation protocol~\citep{GMW}. The
\ourlang AST (and hence the interpreter) does not ``bake in'' standard
\fstar{} constructs like numbers and lists. Rather, inherited language
features appear abstractly in the AST, and their semantics is handled
by a foreign function interface (FFI). This permits \ourlang programs
to take advantage of existing code and libraries available in \fstar{}. 

To prove the interpreter behaves correctly, we prove, in \fstar,
that it correctly implements the formalized distributed semantics.
The circuit library and the GMW implementation
are not verified---while it is possible to verify the circuit
library~\cite{Almeida:2017:FVS:3133956.3134017}, verifying a GMW
implementation is an open research question.
But the stage is set for verified versions to be plugged into
the \ourlang codebase. We characterize the Trusted Computing Base
(TCB) of the \ourlang toolchain in \S\ref{sec:impl}.
  
Using \ourlang we have
implemented several programs, including PSI,
joint median, and a card dealing application
(\S\ref{sec:apps}). For PSI and joint median we
implement two versions: a straightforward one and an optimized one that
improves performance but increases
the number of adversary-observable events. We formally
prove that the optimized and unoptimized
versions are equivalent,
both functionally and w.r.t. privacy of parties' inputs.
Our card dealing application relies on \ourlang's support
for secret shares~\citep{Shamir79}. We formally prove that the card
dealing algorithm always deals a fresh card.

In sum, \ourlang constitutes the first DSL that supports proving
security and correctness properties about \mc programs, which are
executed by a partially verified implementation of a full-featured
language. No prior DSL provides these benefits
(\S\ref{sec:related}). 
The \ourlang implementation, example programs, and
proofs are publicly available on
\iflong
Github at \url{https://github.com/FStarLang/FStar/tree/stratified_last/examples/wysteria}.\footnote{This development was
  done on an older \fstar version, but the core ideas of what we
  present here apply to the present version as well.}
\else
Github.\footnote{This development was
  done on an older \fstar version, but the core ideas of what we
  present here should apply to the present version as well.}
\fi

\section{Verifying and deploying \ourlang programs}
\label{sec:overview}

We illustrate the main concepts of \ourlang by showing, in
several stages, how to program, optimize, and verify the two-party
joint median example~\cite{Kerschbaum11,rastogi13knowledge}.
In this example, two parties, Alice and Bob, each have a set of $n$ distinct,
locally sorted integers, and they want to compute the median of the union of
their sets without revealing
anything else; our running example fixes $n = 2$, for simplicity. 

\subsection{Secure computations with \ls$as_sec$}
\label{sec:assec}
In \ourlang, as in its predecessor Wysteria~\cite{wysteria},
an \mc is written as a single specification that
executes in one of two \emph{computation modes}. The primary mode is called \ls$sec$
mode. In it, a computation is carried out using a \mc protocol
among multiple principals on separate hosts. Here is joint median in \ourlang:
\begin{lstlisting}[xleftmargin=1em, numbers=left, frame=single]
let median a b in_a in_b =
  as_sec {a, b} (fun () -> let cmp = fst (reveal in_a) > fst (reveal in_b) in $\label{line:acomp}$
                        let x3 = if cmp then fst (reveal in_a) else snd (reveal in_a) in $\label{line:acull}$
                        let y3 = if cmp then snd (reveal in_b) else fst (reveal in_a) in $\label{line:bcull}$
                        if x3 > y3 then y3 else x3)
\end{lstlisting}
The four arguments to \ls$median$ are, respectively, principal
identifiers for Alice and Bob, and Alice and Bob's secret inputs expressed as
tuples.
In \ourlang, values specific to each
principal are
\emph{sealed} with the principal's name (which appears in the sealed
container's type). As such, the types of \ls$in_a$ and \ls$in_b$
are, respectively, \ls$sealed {a} (int * int)$ and
\ls$sealed {b} (int * int)$.
The \ls$as_sec ps f$
construct indicates that thunk \ls$f$
should be run in \ls$sec$ mode among principals in the set \ls{ps}.
In this mode, the code has access to the secrets of the principals
\ls$ps$,
which it can reveal using the \ls{reveal} coercion.
As we will see later, the type of \ls{reveal} ensures that
parties cannot \ls{reveal} each others' inputs outside \ls{sec}
mode.\footnote{The runtime
representation of \ls{sealed {a} v} at \ls{b}'s host is an opaque constant
$\bullet$ (\S\ref{sec:deployoverview}).} Also note that the
code freely uses standard \fstar library functions like \ls{fst} and \ls{snd}.
The example extends naturally
to $n > 2$~\cite{10.1007/978-3-540-24676-3_3}.

To run this program, both Alice and Bob would start a \ourlang
interpreter at their host and direct it to run the \ls$median$ function
Upon reaching the
\ls$as_sec$ thunk, the interpreters coordinate with each other to compute the
result using the underlying \mc protocol.
\S\ref{sec:deployoverview} provides more details.



\subsection{Optimizing \ls{median} with \ls$as_par$}

Although \ls$median$ gets the job done, it can
be inefficient for large $n$.
However, it turns out if we reveal the result
of comparison on line~\ref{line:acomp} to both the parties,
then the computation on line~\ref{line:acull} (resp. line~\ref{line:bcull})
can be performed locally
by Alice (resp. Bob) without need of cryptography. Doing so can
massively improve performance:
previous work~\cite{Kerschbaum11} has observed a
$30\times$ speedup for $n = 64$.

This optimized variant is a \emph{mixed-mode} computation, where participants
perform some local computations interleaved with small, jointly
evaluated secure computations. \ourlang's second
computation mode, \ls$par$ mode, supports such mixed-mode computations. The
construct \ls$as_par ps f$ states that each principal in \ls$ps$
should locally execute the thunk \ls$f$, simultaneously; any
principal not in the set \ls$ps$ simply skips the computation. Within
\ls$f$, while running in \ls$par$ mode, principals may engage in
secure computations via \ls$as_sec$. 

Here
is an optimized version of \ls{median} using \ls{as_par}:

\begin{lstlisting}[xleftmargin=1em, numbers=left, frame=single]
let median_opt a b in_a in_b =
 let cmp = as_sec {a, b} (fun () -> fst (reveal in_a) > fst (reveal in_b)) in$\label{line:optacomp}$
 let x3 = as_par {a} (fun () -> if cmp then fst (reveal in_a) else snd (reveal (in_a))) in $\label{line:optacull}$
 let y3 = as_par {b} (fun () -> if cmp then snd (reveal in_b) else fst (reveal (in_b))) in $\label{line:optbcull}$
 as_sec {a, b} (fun () -> if reveal x3 > reveal y3 then reveal y3 else reveal x3) $\label{line:finalm}$
\end{lstlisting}

The secure computation on (line~\ref{line:optacomp}) \emph{only}
computes \ls{cmp} and returns the result to both the parties. Line~\ref{line:optacull}
is then a \ls{par} mode computation involving only Alice
in which she discards one of her inputs based on \ls{cmp}.
Similarly, on line~\ref{line:optbcull}, Bob discards
one of his inputs. Finally, line~\ref{line:finalm} compares the remaining
inputs using \ls{as_sec} and returns the result as the final median.

One might wonder whether \ls$par$ mode is necessary. Could we
program the local parts of a mixed-mode program in normal \fstar, and
use a special compiler to convert the \ls$sec$ mode parts to circuits
and pass them to a GMW \mc service? We could, but it would complicate
both writing \mc{s} and formally reasoning that the whole computation is
correct and secure.
In particular, programmers would need to write one program for each
party that performs a different local computation (as in
\ls{median_opt}). The potential interleaving among local computations
and their synchronization behavior when securely computing together
would be a source of possible error and thus must be considered in any
proof. For example, Alice's code might have a bug in it that prevents
it from reaching a synchronization point with Bob, to do a GMW-based
\mc. For \ourlang, the situation is much simpler. Programmers may
write and maintain a single program. This program can be formally
reasoned about directly using a SIMD-style, ``single-threaded''
semantics, per the soundness result from \S\ref{sec:metatheory}. This
semantics permits reasoning about the coordinated behavior of multiple
principals, without worry about the effects of interleavings or wrong
synchronizations. Thanks to \ls{par} mode, invariants about
coordinated local computations are directly evident since we can
soundly assume about lockstep behavior (e.g., loop iterations in the
PSI example in \S\ref{sec:apps}).


\subsection{Embedding a type system for \ourlang in \fstar}
\label{sec:wysts}

Designing high-level, multi-party computations is relatively
easy using Wysteria's abstractions. Before trying to run such a
computation, we might wonder:  

\begin{enumerate}
\item Is it \emph{realizable}? For example, does a computation
that is claimed to be executed only by some principals \ls$ps$ (e.g.,
using an \ls$as_par ps$ or an \ls$as_sec ps$) only ever access data
belonging to \ls$ps$?

\item Is it \emph{correct?} For example, does \ls$median_opt$
  correctly compute the median of Alice and Bob's inputs?

\item Is it \emph{secure}? For example, do the optimizations in
  \ls$median_opt$, which produce more visible outputs, potentially
  leak more about the inputs?
\end{enumerate}

By embedding \ourlang in \fstar and leveraging its extensible,
monadic, dependent type-and-effect system, we address each of these
three questions.
We define a new indexed monad called \ls$Wys$ for computations that
use \mc combinators \ls$as_sec$ and \ls$as_par$. Using \ls$Wys$
along with the \ls$sealed$ type, we can ensure that protocols are
realizable. Using \fstar's capabilities for formal verification, we
can reason about a computation's correctness. By characterizing 
observable events as part of \ls$Wys$, we can define trace
properties of \mc programs, to reason about security.

To elaborate on the last: we are interested in \emph{application-level} security
properties, assuming that the underlying 
cryptographic \mc protocol (GMW~\cite{GMW} in our implementation) is secure.
In particular,
the \ls{Wys} monad models the \emph{ideal} behavior of \ls{sec} mode---a
secure computation reveals only the final output and nothing
else. Thus the programmer could reason, for example, that optimized
\mc programs reveal no more than their unoptimized versions.
To relate the proofs over ideal functionality to the actual
implementation, as is standard, we rely on the security of the
cryptographic protocol and the composition theorem~\cite{Canetti:2000:SCM:2724987.2725177}
to postulate that the implementation securely realizes the ideal
specification.

\Paragraph{The \ls$Wys$ monad} The \ls$Wys$ monad provides several features. First, all DSL code is
typed in this monad, encapsulating it from the rest of \fstar. Within
the monad, computations and their specifications can make use of two
kinds of \emph{ghost state}: \emph{modes} and \emph{traces}.
The mode of a computation indicates whether the computation is running
in an \ls$as_par$ or in an \ls$as_sec$ context.
The trace of a computation records the sequence and nesting structure
of  outputs of the jointly executed \ls$as_sec$ expressions---the result of a
computation and its trace constitute its observable behavior.
The \ls$Wys$ monad is, in essence, the product of a reader monad on
modes and a writer monad on traces~\cite{Wadler:1995:MFP:647698.734146,Moggi:1991:NCM:116981.116984}.

Formally, we define the following \fstar types for modes and traces. A
mode \ls$Mode m ps$ is a pair of a mode tag (either \ls$Par$
or \ls$Sec$) and a set of principals \ls$ps$. A \ls$trace$ is a forest
of trace element (\ls$telt$) trees. The leaves of the trees record
messages \ls$TMsg x$ that are received as the result of executing
an \ls$as_sec$ thunk. The tree structure represented by the
\ls$TScope ps t$ nodes record the set of principals that are able to
observe the messages in the trace \ls$t$.

\begin{lstlisting}
type mtag = Par | Sec
type mode = Mode: m:mtag -> ps:prins -> mode
type telt = | TMsg  : x:$\alpha$ -> telt | TScope: ps:prins -> t:list telt -> telt
type trace = list telt
\end{lstlisting}

Every \ourlang computation $e$ has a monadic computation
type \ls$Wys t pre post$.
The type indicates that $e$ is in the \ls$Wys$ monad (so it may
perform multi-party computations);
\ls$t$ is its result type;
\ls$pre$ is a pre-condition on the mode in which \ls$e$ may be executed;
and \ls$post$ is a post-condition relating the computation's mode, its
result value, and its trace of observable events.
When run in a context with mode \ls$m$ satisfying the pre-condition
predicate \ls$pre m$, $e$ may produce the trace \ls$tr$,
and if and when it returns, the result is
a \ls$t$-typed value \ls$v$ validating \ls@post m v tr@.
The style of indexing a monad with a computation's pre- and
post-condition is a standard
technique~\citep{atkey09parameterised,nmb08htt,fstar2016}---we defer
the definition of the monad's \ls$bind$ and \ls$return$ to the
actual implementation and focus instead on specifications of \ourlang
specific combinators.
We describe
\ls$as_sec$, \ls$reveal$, and \ls$as_par$, and how we give them types in
\fstar, leaving the rest to
Figure~\ref{fig:wysstar-api} in the Appendix.
By convention, any free variables in the type signatures are
universally prenex quantified.

\Paragraph{Defining \ls$as_sec$ in \ourlang} 
~

\begin{lstlisting}[numbers=left]
val as_sec: ps:prins -> f:(unit -> Wys a pre post) -> Wys a
  (requires (fun m     -> m=Mode Par ps /\ pre (Mode Sec ps))) $\label{line:assec-requires}$
  (ensures  (fun m r tr -> tr=[TMsg r] /\ exists t. post (Mode Sec ps) r t)))$\label{line:assec-ensures}$
\end{lstlisting}

The type of \ls$as_sec$ is \emph{dependent} on the first
parameter, \ls$ps$. Its second argument \ls$f$ is the thunk to be
evaluated in \ls$sec$ mode. The result's computation type has the
form
\ls@Wys a (requires $\phi$) (ensures $\psi$)@, for some pre-condition
and post-condition predicates $\phi$ and $\psi$, respectively. 
We use the
\ls$requires$ and \ls$ensures$ keywords for readability---they are not semantically significant.

The pre-condition of \ls$as_sec$ is a predicate on the mode \ls$m$ of
the computation in whose context
\ls$as_sec ps f$ is called.
For all the \ls$ps$ to jointly execute \ls$f$, we require all
of them to transition to perform the \ls$as_sec ps f$ call
simultaneously, i.e., the current mode must be
\ls$Mode Par ps$.
We also require the pre-condition \ls$pre$
of \ls$f$ to be valid once the mode has transitioned to
\ls$Mode Sec ps$---line~\ref{line:assec-requires} says just this.

The post-condition of \ls$as_sec$ is a predicate
relating the initial mode \ls$m$, the result \ls$r:a$, and the
trace \ls$tr$ of the computation.
Line~\ref{line:assec-ensures} states that the trace of a secure
computation \ls$as_sec ps f$ is just a singleton \ls$[TMsg r]$,
reflecting that its execution reveals only result
\ls$r$.
Additionally, it ensures that the result \ls$r$ is related to the mode
in which \ls$f$ is run (\ls$Mode Sec ps$) and some trace \ls$t$
according to \ls$post$, the post-condition of \ls$f$. The API models the
``ideal functionality'' of secure computation protocols (such as GMW)
where the participants only observe the final result.

\Paragraph{Defining \ls$reveal$ in \ourlang} As discussed earlier, a
value \ls$v$ of type \ls$sealed ps t$ encapsulates a \ls$t$ value that
can be accessed by calling \ls$reveal v$. This call should only
succeed under certain circumstances. For example, in \ls$par$ mode, Bob
should not be able to reveal a value of type \ls$sealed {Alice} int$.
The type of \ls$reveal$ makes the access control rules clear:





\begin{lstlisting}
val unseal: sealed ps $\alpha$ -> Ghost $\alpha$

val reveal: x:sealed ps $\alpha$ -> Wys $\alpha$
  (requires (fun m -> m.mode=Par ==> m.ps $\subseteq$ ps /\ m.mode=Sec ==> m.ps $\cap$ ps $\neq$ $\emptyset$))
  (ensures (fun m r tr -> r=unseal x /\ tr=[]))
\end{lstlisting}

The \ls{unseal} function is a \ls{Ghost} function, meaning that it can
only be used in specifications for reasoning purposes. On the other
hand, \ls{reveal} can be called in the concrete \ourlang programs.
%
Its precondition says that when executing in \ls$Mode Par ps'$,
\emph{all} current participants must be listed in the seal, i.e.,
\ls@ps' $\subseteq$ ps@.
However, when executing in \ls$Mode Sec ps'$, only a subset of
current participants is required:
\ls@ps' $\cap$ ps $\neq$ $\emptyset$@.
This is because the secure computation is executed jointly by all of
\ls$ps'$, so it can access any of their individual data. 
The postcondition of \ls{reveal} relates the result \ls{r} to the
argument \ls{x} using the \ls{unseal} function.
%

\Paragraph{Defining \ls$as_par$ in \ourlang}
~

\begin{lstlisting}[numbers=left]
val as_par: ps:prins -> (unit -> Wys a pre post) -> Wys (sealed ps a)
  (requires (fun m     -> m.mode=Par    /\ ps $\subseteq$ m.ps /\ can_seal ps a /\ pre (Mode Par ps)))
  (ensures  (fun m r tr -> exists t. tr=[TScope ps t] /\ post (Mode Par ps) (unseal r) t)))
\end{lstlisting}

The type of \ls{as_par} enforces the current mode to be \ls{Par},
and \ls{ps} to be a subset of current principals. Importantly,
the API scopes the trace \ls{t} of \ls{f} to model the fact that
any observables of \ls{f} are only visible to the
principals in \ls{ps}. Note that \ls{as_sec} did not require such scoping,
as there \ls{ps} and the set of current principals in \ls{m} are the same.

\subsection{Correctness and security verification}
\label{sec:verification}

Using the \ls$Wys$ monad and the \ls$sealed$ type, we can write down
precise types for our \ls$median$ and \ls$median_opt$ programs,
proving various useful properties. We discuss
the statements of the main lemmas and the overall proof structure.
By programming the protocols as a single specification using the high-level
abstractions provided by \ourlang, our proofs are relatively straightforward---in
all the proofs of this section, \fstar required no additional hints.
In particular, we rely heavily on the view that both parties execute
(different fragments of) the same code, thus avoiding the unwieldy
task of reasoning about low-level message passing.

\Paragraph{Correctness and security of \ls{median}} We first define a pure specification of median of
two \ls{int} tuples:

\begin{lstlisting}
let median_of (x1, x2) (y1, y2) = let (_, m, _, _) = sort x1 x2 y1 y2 in m
\end{lstlisting}

Further, we capture the preconditions using the following predicate:

\begin{lstlisting}
let median_pre (x1, x2) (y1, y2) = x1 < x2 $\wedge$ y1 < y2 $\wedge$ distinct x1 x2 y1 y2
\end{lstlisting}

Using these, we prove the following top-level specification for \ls{median}:

\begin{lstlisting}
val median: in_a:sealed {a} (int * int) -> in_b:sealed {b} (int * int) -> Wys int
  (requires (fun m -> m = Mode Par {a, b}))  (* should be called in the Par mode *)
  (ensures (fun m r tr -> let in_a, in_b = unseal in_a, unseal in_b in
      (median_pre in_a in_b ==> r = median_of in_a in_b) /\    (* functional correctness *)
      tr = [TMsg ra]))    (* trace is just the final value *)
\end{lstlisting}

This signature establishes that when Alice and Bob
simultaneously execute \ls$median$ (in \ls$Par$
mode), with secrets \ls$in_a$ and \ls$in_b$, then if and when the
protocol terminates,
(a) if their inputs satisfy the
precondition \ls{median_pre}, then the result is the joint median of
their inputs and (b) the observable trace consists
only of the final result, as there is but  a single \ls{as_sec} thunk
in \ls{median}, i.e., it is \emph{secure}.

\Paragraph{Correctness and security of \ls{median_opt}}
The security proof of \ls{median_opt} is particularly interesting, because 
the program intentionally reveals more than just the final result,
i.e., the output of the first comparison.
We would like to verify that this additional information does not
compromise the privacy of the parties' inputs.
To do this, we take the following approach.

First, we characterize the observable trace of \ls{median_opt} as a pure,
specification-only function. Then, using relational reasoning, we prove
a \emph{noninteference with delimited release} property~\citep{sm03delimited}
on these traces. Essentially we prove that, for two runs of \ls{median_opt}
where Bob's inputs and the output median are the same, the observable traces
are also same irrespective of Alice's inputs. Thus, from Alice's perspective, the
observable trace does not reveal more to Bob than what the output already does.
We prove this property symmetrically for Bob.

We start by defining a trace function for \ls{median_opt}:

\begin{lstlisting}
let opt_trace a b (x1, _) (y1, _) m = [
  TMsg (x1 > y1);  (* observable from the first as_sec *)
  TScope {a} []; TScope {b} [];  (* observables from two local as_par *)
  TMsg m ]  (* observable from the final as_sec *)
\end{lstlisting}

\noindent A trace will have four elements: output of the first \ls{as_sec}
computation, two empty scoped traces for the two local \ls{as_par}
computations, and the final output.

Using this function, we prove correctness of \ls{median_opt}, thus:

\begin{lstlisting}
val median_opt: in_a:sealed {a} (int * int) -> in_b:sealed {b} (int * int) -> Wys int
  (requires (fun m -> m = Mode Par {a, b}))  (* should be called in the Par mode *)
  (ensures (fun m r tr -> let in_a = unseal in_a in let in_b = unseal in_b in
    (median_pre in_a in_b ==> r = median_of in_a in_b) /\  (* functional correctness *)
    tr = opt_trace a b in_a in_b m  (* opt_trace precisely describes the observable trace *)

\end{lstlisting}

The delimited release property is then captured by the following lemma:

\begin{lstlisting}
val median_opt_is_secure_for_alice: a:prin -> b:prin
  -> in_a$_1$:(int * int) -> in_a$_2$:(int * int) ->  in_b:(int * int) (* possibly diff a1, a2 *)
  -> Lemma (requires (median_pre in_a$_1$ in_b /\ median_pre in_a$_2$ in_b /\ 
                      median_of in_a$_1$ in_b = median_of in_a$_2$ in_b)) (* but same median *)
            (ensures  (opt_trace a b in_a$_1$ in_b (median_of in_a$_1$ in_b) = (* ensures .. *)
                     opt_trace a b in_a$_2$ in_b (median_of in_a$_2$ in_b))) (* .. same trace *)
\end{lstlisting}

The lemma proves that for two runs of \ls$median_opt$ where Bob's
input and the final output remain same, but Alice's inputs vary
arbitrarily, the observable traces are the same. As such, no more
information about information leaks about Alice's inputs via the traces than
what is already revealed by the output. We
also prove a symmetrical lemma \ls{median_opt_is_secure_for_bob}.

In short, because the \ls{Wys} monad provides programmers
with the observable traces in the logic, they can then be used to prove
properties, relational or otherwise, in the pure fragment of
\fstar outside the \ls{Wys} monad. We present more examples and their
verification details in \S\ref{sec:apps}.

\subsection{Deploying \ourlang programs}
\label{sec:deployoverview}

\begin{figure}[t]
\centering
  \includegraphics[scale=0.55]{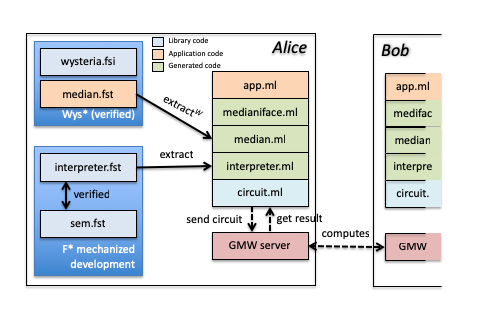}
  \caption{Architecture of an \ourlang deployment}
  \label{fig:distarch}
\end{figure}

Having defined a proved-secure \mc program in \ourlang, how do we run it?
Doing so requires the following steps (Figure~\ref{fig:distarch}).
First, we run the
\fstar compiler in a special mode that \emph{extracts} the \ourlang
code (say \ls{psi.fst}),
into the \ourlang AST as a data structure (in \ls{psi.ml}).
Except for the \ourlang specific nodes (\ls{as_sec}, \ls{as_par}, etc.),
the rest of the program is extracted
into \emph{FFI nodes} that indicate the use of, or calls into,
functionality provided by \fstar itself.

The next step is for each party to run the extracted
AST using the \ourlang interpreter. This interpreter is written in
\fstar and we have proved (see \S\ref{sec:impl}) it implements a deep embedding of the \ourlang
semantics, also specified in \fstar (Figures~\ref{fig:dsl-proto-semantics} and \ref{fig:dsl-tgt-semantics},
\S\ref{sec:formal}). The
interpreter is extracted to OCaml by the usual \fstar extraction.
Each party's interpreter executes the AST locally
until it reaches an \ls$as_sec ps f$ node, 
where the interpreter's back-end compiles
\ls$f$, on-the-fly, for particular values of the secrets in \ls$f$'s
environment, to a boolean circuit. First-order, loop-free code can be
compiled to a circuit; \ourlang provides specialized support for
several common combinators
(e.g., \ls{fst}, \ls{snd}, list combinators such as \ls$List.intersect$, \ls$List.mem$,
\ls$List.nth$ etc.).

The circuit is handed to a library by Choi et al.~\citet{cryptoeprint:2011:257}
that implements the GMW~\citep{GMW} \mc protocol.
Running the GMW protocol involves the parties in \ls{ps}
generating and communicating (XOR-based) secret shares \citep{Shamir79} for
their secret inputs, and then cooperatively evaluating the boolean
circuit for \ls$f$ over them.

One obvious question is how both parties are able to get this process
off the ground, given that they don't know some of the inputs (e.g., other
parties' secrets).
The \ls{sealed} abstraction helps here.
Recall that for \ls$median$,
the types of the inputs are of the form \ls$sealed {a} (int * int)$ and
\ls$sealed {b} (int * int)$. When the program is run on Alice's host,
the former will be a pair of Alice's values, whereas the latter
will be an opaque constant (which we denote as $\bullet$). The
reverse will be true on Bob's host. When the circuit is constructed, 
each principal links their non-opaque inputs to the relevant input
wires of the circuit. Similarly, the output map component of each party
is derived from their output wires in the circuit, and thus,
each party only gets to see their own output.

\section{Formalizing and implementing \ourlang}
\label{sec:formal}








\newcommand{\ext}[1]{\ensuremath{\mathsf{#1}}}
\newcommand{\ps}{\ensuremath{s}}

\begin{figure}[t]
\[
\begin{array}{@{}@{}rlcl}
\text{Principal} \quad p & & & \text{        Principal set} \quad \ps \quad \text{                    FFI const} \quad \ext{c}, \ext{f}\\
\text{Constant} & c & ::= & p \mid \ps \mid () \mid \kw{true} \mid \kw{false} \mid \ext{c}\\
\text{Expression} & e & ::=
& \kw{as_par}\;e_1\;e_2 \mid \kw{as_sec}\;e_1\;e_2 \mid \kw{seal}\;e_1\;e_2 \mid \kw{reveal}\;e \mid \kw{ffi}\;\ext{f}\;\bar{e}\\
& & \mid & \kw{mkmap}\;e_1\;e_2 \mid \kw{project}\;e_1\;e_2 \mid \kw{concat}\;e_1\;e_2\\
& & \mid & c \mid x \mid \kw{let}\;x = e_1\;\kw{in}\;e_2 \mid \lambda x.e \mid e_1\;e_2 \mid \kw{fix}\;f.\lambda x.e \mid \kw{if}\;e_1\;\kw{then}\;e_2\;\kw{else}\;e_3\\
\end{array}
\]
\caption{\ourlang syntax}
\label{fig:dsl-syntax}
\end{figure}


In the previous section, we presented examples of verifying properties
about \ourlang programs using \fstar{}'s logic. However,
these programs are not executed using the \fstar{} (single-threaded) semantics; they
have a distributed semantics involving multiple parties. So,
how do the properties that we verify using \fstar{} carry over?

In this section, we present the metatheory that answers this
question. First, we formalize the \ourlang single-threaded (ST) semantics,
that
faithfully models the \fstar{} semantics of the \ourlang API presented
in \S\ref{sec:overview}. Next, we formalize
the distributed (DS) semantics that multiple parties use 
to run \ourlang programs. Then we prove the former is \emph{sound}
with respect to the latter, so that properties proved of programs
under ST apply when run under DS\@. We have mechanized the proof of
this theorem in \fstar{}.  


\subsection{Syntax}

Figure~\ref{fig:dsl-syntax} shows the complete
syntax of \ourlang. Principals and principal sets are first-class
values, and are denoted by $p$ and $\ps$ respectively. Constants in
the language also include $()$ (unit), booleans, and FFI
constants \ext{c}. 
Expressions $e$ include the regular forms for functions, applications,
let bindings, etc. and the \ourlang-specific constructs. Among the
ones that we have not seen in \S\ref{sec:overview}, expression
$\kw{mkmap}\;e_1\;e_2$ creates a map from principals in $e_1$
(which is a principal set) to the value computed by
$e_2$. $\kw{project}\;e_1\;e_2$ projects the value of principal $e_1$
from the map $e_2$, and $\kw{concat}\;e_1\;e_2$ concatenates the two
maps. The maps are used if an \ls{as_sec} computation returns
different outputs to the parties.


Host language (i.e., \fstar) constructs are also part of the syntax of
\ourlang, including constants \ext{c} include strings, integers, lists,
tuples, etc. Likewise, host language functions/primitives can be called
from \ourlang---$\kw{ffi}\;\ext{f}\;\bar{e}$ is the invocation of a
host-language function \ext{f} with arguments
$\bar{e}$. The FFI confers two benefits. First, it simplifies the core
language while still allowing full consideration of security relevant
properties. Second, it helps the language scale by incorporating
many of the standard features, libraries, etc. from the host language.





\begin{figure}[t]
\[
\begin{array}{@{}rlcl}
\text{Map} & m & ::= & \cdot \mid m[p \mapsto v]\\
\text{Value} & v & ::= & p \mid \ps \mid () \mid \kw{true} \mid \kw{false} \mid \kw{sealed}\;\ps\;v \mid m \mid \ext{v} \mid (L, \lambda x.e) \mid (L, \kw{fix}\;f.\lambda x.e) \mid \bullet\\
\text{Mode} & M & ::= & \kw{Par}\;\ps \mid \kw{Sec}\;\ps\\
\text{Context} & E & ::= & \langle\rangle \mid \kw{as_par}\;\langle\rangle\;e \mid \kw{as_par}\;v\;\langle\rangle \mid \kw{as_sec}\;\langle\rangle\;e \mid \kw{as_sec}\;v\;\langle\rangle \mid \ldots\\
\text{Frame} & F & ::= & (M, L, E, T)\\
\text{Stack} & X & ::= & \cdot \mid F,X\\
\text{Environment} & L & ::= & \cdot \mid L[x \mapsto v]\\
\text{Trace element} & t & ::= & \kw{TMsg}\;v \mid \kw{TScope}\;\ps\;T\\
\text{Trace} & T & ::= & \cdot \mid t, T\\
\text{Configuration} & C & ::= & M; X; L; T; e\\\\
\text{Par component} & P & ::= & \cdot \mid P[p \mapsto C]\\
\text{Sec component} & S & ::= & \cdot \mid S[\ps \mapsto C]\\
\text{Protocol} & \pi & ::= & P; S\\
\end{array}
\]
\caption{Runtime configuration syntax} 
\label{fig:dsl-runtime-syntax}
\end{figure}

\begin{figure*}[t]
\[
\begin{array}{l}
\inferrule*[lab=S-aspar]
{
e_1 = \kw{as_par}\;\ps\;(L_1, \lambda x.e) \quad M = \kw{Par}\;\ps_1 \quad \ps \subseteq \ps_1 \\\\
X_1 = (M; L; \kw{seal}\;\ps\;\langle\rangle; T), X
}
{
M; X; L; T; e_1 \rightarrow \kw{Par}\;\ps;X_1; L_1[x \mapsto ()]; \cdot; e
}
\hspace{0.3cm}
\inferrule*[lab=S-parret]
{
X = (M_1; L_1; \kw{seal}\;\ps\;\langle\rangle; T_1), X_1\\\\
\kw{can_seal}\;\ps\;v \quad
T_2 = \kw{append}\;T_1\;[\kw{TScope}\;\ps\;T]
}
{
M; X; L; T; v \rightarrow M_1; X_1; L_1; T_2; \kw{sealed}\;\ps\;v
}
\\\\
\inferrule*[lab=S-assec]
{
e_1 = \kw{as_sec}\;\ps\;(L_1, \lambda x.e) \quad M = \kw{Par}\;\ps \\\\
X_1 = (M; L; \langle\rangle\; T), X
}
{
M; X; L; T; e_1  \rightarrow \kw{Sec}\;\ps; X_1; L_1[x \mapsto ()]; \cdot; e
}
\hspace{0.9cm}
\inferrule*[lab=S-secret]
{
\kw{is_sec}\;M \quad X = (M_1; L_1; \langle\rangle; T), X_1 \\\\
T_1 = \kw{append}\;T\;[\kw{TMsg}\;v] \quad
}
{
M; X; L; \cdot; v \rightarrow M_1; X_1; L_1; T_1; v
}
\end{array}
\]
\caption{\ourlang ST semantics (selected rules)}
\label{fig:src-semantics}
\end{figure*}

\subsection{Single-threaded semantics}
\label{sec:st}
We formalize the semantics in the style of Hieb and
Felleisen~\cite{felleisen1992revised}, where the redex is chosen by
(standard, not shown) \emph{evaluation contexts} $E$, which prescribe 
left-to-right, call-by-value evaluation order.
The ST semantics, a model of the \fstar{} semantics and the \ourlang API,
defines a judgment
$C \rightarrow C'$ that represents a single step of an abstract
machine (Figure~\ref{fig:src-semantics}). Here, $C$ is a \emph{configuration} $M; X; L; T; e$. This
five-tuple consists of a mode $M$, a stack $X$, a local environment
$L$, a trace $T$, and an expression $e$. The syntax for these elements
is given in Figure~\ref{fig:dsl-runtime-syntax}. The value form
$\ext{v}$ represents the host language (FFI) values. The stack and
environment are standard; trace $T$ and mode $M$ were discussed in
the previous section. 

For space reasons, we focus on the two main \ourlang constructs \kw{as_par}
and \kw{as_sec}. 
Appendix B shows rules for other \ourlang specific constructs.

Rules {\sc{S-aspar}} and {\sc{S-parret}} (Figure~\ref{fig:src-semantics})
reduce an \kw{as_par}
expression once its arguments are fully evaluated---its first argument $s$
is a principal set, while the second argument $(L_1, \lambda x.e)$ is a closure
where $L_1$ captures the free variables of thunk $\lambda x.e$. {\sc{S-aspar}}
first checks that the current mode $M$ is \kw{Par} and contains all the
principals from the set $\ps$. It then pushes a
$\kw{seal}\;\ps\;\langle\rangle$ frame on the stack, and starts evaluating
$e$ under the environment $L_1[x \mapsto ()]$. The rule {\sc{S-asparret}} pops the frame and
seals the result,
so that it is accessible only to the
principals in $\ps$. The rule also creates a trace element
$\kw{TScope}\;\ps\;T$, essentially making observations during the
reduction of $e$ (i.e., $T$) visible only to principals in $\ps$.

Turning to \kw{as_sec}, the rule {\sc{S-assec}} checks the precondition of the API,
and the rule {\sc{S-assecret}} generates a trace observation $\kw{TMsg}\;v$,
as per the postcondition of the API. As mentioned before, \ls{as_sec} semantics
models the ideal, trusted third-party semantics of secure computations where the
participants only observe the final output. We
can confirm that the rules implement the types of \kw{as_par} and
\kw{as_sec} shown in \S\ref{sec:overview}.

\begin{figure*}[t]
\[
\begin{array}{l}
\inferrule*[lab=P-par]
{
C \leadsto C'
}
{
P[p \mapsto C]; S \longrightarrow P[p \mapsto C']; S
}
\hspace{0.6cm}
\inferrule*[right=P-enter]
{
\forall p \in \ps.\;P[p].e = \kw{as_sec}\;\ps\;(L_p, \lambda x.e) \\\\
\ps \not\in \kw{dom}(S) \quad
L = \kw{combine}\;\bar{L}_p \quad
}
{
P; S \longrightarrow P; S[\ps \mapsto \kw{Sec}\;\ps; \cdot; L[x \mapsto ()]; \cdot; e]
}
\\\\
\inferrule*[lab=P-sec]
{
C \rightarrow C'
}
{
P; S[\ps \mapsto C] \longrightarrow P; S[\ps \mapsto C']
}
\hspace{0.6cm}
\inferrule*[lab=P-exit]
{
S[\ps] = \kw{Sec}\;\ps; \cdot; L; T; v \\\\
P' = \forall p \in \ps.\;P[p \mapsto P[p] \triangleleft (\kw{slice_v}\;p\;v)] \quad
S' = S \setminus \ps
}
{
P; S \longrightarrow P'; S'
}
\end{array}
\]
\caption{Distributed semantics, multi-party rules}
\label{fig:dsl-proto-semantics}
\end{figure*}

\begin{figure*}[t]
\[
\begin{array}{l}
\inferrule*[lab=L-aspar1]
{
e_1 = \kw{as_par}\;\ps\;(L_1, \lambda x.e) \quad p \in \ps \\\\
X_1 = (M; L; \kw{seal}\;\ps\;\langle\rangle; T), X
}
{
\kw{Par}\;p; X; L; T; e_1 \leadsto \kw{Par}\;p; X_1; L_1[x \mapsto ()]; \cdot; e
}
\hspace{0.3cm}
\inferrule*[lab=L-parret]
{
X = (M; L_1; \kw{seal}\;\ps\;\langle\rangle; T_1), X_1 \\\\
T_2 = \kw{append}\;T_1\;T \quad v_1 = \kw{sealed}\;\ps\;v
}
{
\kw{Par}\;p; X; L; T; v \leadsto \kw{Par}\;p; X_1; L_1; T_2; v_1
}
\\\\
\hspace{1cm}\inferrule*[left=L-aspar2]
{
p \not\in \ps
}
{
\kw{Par}\;p; X; L; T; \kw{as_par}\;\ps\;(L_1, \lambda x.e) \leadsto \kw{Par}\;p; X; L; T; \kw{sealed}\;\ps\;\bullet
}
\end{array}
\]
\caption{Distributed semantics, selected local rules {\small{(the mode $M$ is always \kw{Par}\;$p$)}}}
\label{fig:dsl-tgt-semantics}
\end{figure*}

\subsection{Distributed semantics}
\label{sec:ds}
In the DS semantics, principals
evaluate the same program locally and
asynchronously until they reach a secure computation, at which point
they synchronize to jointly perform the computation.
The semantics consists of two parts: (a) a judgment of the form
$\pi \longrightarrow \pi'$ (Figure~\ref{fig:dsl-proto-semantics}),
where a protocol $\pi$ is a tuple $(P; S)$
such that $P$ maps each principal to its local
configuration and $S$ maps a set of principals to the configuration of
an ongoing, secure computation; and (b) a local evaluation judgment
$C \leadsto C'$ (Figure~\ref{fig:dsl-tgt-semantics}) to model how a single
principal behaves while in \ls$par$ mode.

Rule {\sc{P-Par}} in Figure~\ref{fig:dsl-proto-semantics} models a
single party taking a step, per the local evaluation rules.
Figure~\ref{fig:dsl-tgt-semantics} shows these rules
for \ls{as_par}. 
(See Appendix B for more local evaluation rules.)
A principal either participates in the
\kw{as_par} computation, or skips it. Rules {\sc{L-aspar1}} and {\sc{L-parret}}
handle the case when $p \in \ps$, and so, the principal $p$
participates in the computation. The rules closely mirror the
corresponding ST semantics rules in Figure~\ref{fig:src-semantics}. One difference in the rule
{\sc{L-asparret}} is that the trace $T$ is not scoped. In the DS
semantics, traces only contain \kw{TMsg} elements; i.e., a trace is
the (flat) list of secure computation outputs observed by that
active principal. 
If $p \not\in \ps$, then the principal skips the computation with the
result being a sealed value containing the opaque constant $\bullet$ (rule
{\sc{L-aspar2}}). The contents of the sealed value do not matter,
since the principal will not be allowed to unseal the value anyway.

As should be the case, there
are no local rules for \kw{as_sec}---to perform a
secure computation parties need to combine their data and jointly do
the computation. Rule {\sc{P-enter}} in Figure~\ref{fig:dsl-proto-semantics}
handles the case when principals enter a secure
computation. It requires that all
the principals $p \in \ps$ must have the expression form
$\kw{as_sec}\;\ps\;(L_p, \lambda x.e)$, where $L_p$ is their local
environment associated with the closure. Each party's local
environment contains its secret values (in addition to some public
values). Conceptually, a secure computation \emph{combines} these
environments, thereby producing a joint view, and evaluates $e$ under
the combination. We define an auxiliary \ls$combine$ function for this
purpose:

\begin{lstlisting}
combine_v ($\bullet$, v) = v
combine_v (v, $\bullet$) = v
combine_v (sealed s v$_1$, sealed s v$_2$) = sealed s (combine_v v$_1$ v$_2$)
...
\end{lstlisting}

The rule {\sc{P-enter}} combines the principals' environments, and
creates a new entry in the $S$ map. The principals are now waiting for
the secure computation to finish. Rule {\sc{P-sec}} models a stepping rule
inside the \ls{sec} mode.

The rule {\sc{P-exit}} applies when a secure computation has
completed and returns results to the waiting principals. If the
secure computation terminates with value $v$, each principal $p$ gets
the value $\kw{slice_v}\;p\;v$. The \kw{slice_v} function is
analogous to \kw{combine}, but in the opposite direction---it
strips off the parts of $v$ that are not accessible to $p$:

\begin{lstlisting}
slice_v p (sealed s v) = sealed s $\bullet$, if p $\not\in$ s
slice_v p (sealed s v) = sealed s (slice_v p v), if p $\in$ s
...
\end{lstlisting}

In the rule {\sc{P-exit}}, the $\triangleleft$ notation is defined as:

\vspace{0.1cm}
$M; X; L; T; \_\;\triangleleft\;v = M; X; L; \kw{append}\;T\;[\kw{TMsg}\;v]; v$
\vspace{0.1cm}

That is, the returned value is also added to the principal's trace to
note their observation of the value.

\subsection{Metatheory}
\label{sec:metatheory}

Our goal is to show that the ST semantics faithfully represents the
semantics of \ourlang programs as they are executed by multiple parties, i.e.,
according to the DS semantics. We do this by proving
\emph{simulation} of the ST semantics by the DS semantics, and by
proving \emph{confluence} of the DS semantics. Our \fstar{}
development mechanizes all the metatheory presented in this section.

\Paragraph{Simulation} We define a \kw{slice}\;\ps\;$C$
function that returns the corresponding protocol $\pi_C$ for an ST
configuration $C$. In the $P$ component of $\pi_C$, each
principal $p \in \ps$ is mapped to their \emph{slice} of the
protocol. For slicing values, we use the same \kw{slice_v}
function as before. Traces are sliced as follows:

\begin{lstlisting}
slice_tr p (TMsg v) = [TMsg (slice_v p v)]
slice_tr p (TScope s T) = slice_tr p T,  if p $\in$ s
slice_tr p (TScope s T) = [],  if p $\not\in$ s
\end{lstlisting}

The slice of an expression (e.g., the source program) is itself. For
all other components of $C$, slice functions are defined analogously.

We say that $C$ is \emph{terminal} if it is in \kw{Par} mode and is fully
reduced to a value
(i.e. when $C = \_; X; \_; \_; e$,  $e$ is a value and $X$ is empty). Similarly, a protocol
$\pi = (P, S)$ is terminal if $S$ is empty and all the local
configurations in $P$ are terminal. The simulation theorem
is then the following:

\begin{theorem}[Simulation of ST by DS]
Let $\ps$ be the set of all principals. If $C_1 \rightarrow^{*} C_2$,
and $C_2$ is terminal, then there exists some derivation
$(\kw{slice}\;\ps\;C_1) \longrightarrow^{*} (\kw{slice}\;\ps\;C_2)$
such that
$(\kw{slice}\;\ps\;C_2)$ is terminal.
\end{theorem}


To state \emph{confluence}, we first
define the notion of \emph{strong termination}.

\begin{definition}[Strong termination]
If all possible runs of protocol $\pi$ terminate at $\pi_t$, we say
$\pi$ \emph{strongly terminates in $\pi_t$}, written $\pi \Downarrow
\pi_t$. 
\end{definition}

Our confluence result then says:

\begin{theorem}[Confluence of DS]
If $\pi \longrightarrow^{*} \pi_t$ and $\pi_t$ is terminal, then
$\pi \Downarrow \pi_t$.
\end{theorem}

Combining the two theorems, we get a corollary that establishes the
soundness of the ST semantics w.r.t. the DS semantics:

\begin{corollary}[Soundness of ST semantics]
Let $\ps$ be the set of all principals. If $C_1 \rightarrow^* C_2$,
and $C_2$ is terminal, then 
$(\kw{slice}\;\ps\;C_1) \Downarrow (\kw{slice}\;\ps\;C_2)$.
\end{corollary}

Now suppose that for a \ourlang source program, we prove in \fstar{}
a post-condition that the result is \kw{sealed} \kw{alice} $n$, for
some $n > 0$. By the soundness of the ST semantics, we can conclude
that when the program is run in the DS semantics, it may diverge, but
if it terminates, \kw{alice}'s output will also
be \kw{sealed} \kw{alice} $n$, and for all other principals their
outputs will be \kw{sealed} \kw{alice} $\bullet$. Aside from the
correspondence on results, our semantics also covers correspondence
on traces. Thus the
correctness and security properties that we prove about a \ourlang
program using \fstar's logic, hold for the program that actually runs.

\subsection{Implementation}
\label{sec:impl}

The formal semantics presented in the prior section is
mechanized as 
an inductive type in \fstar. This style is useful for
proving properties, but does not directly translate to an
implementation. Therefore, we implement an interpretation function
\ls$step$ in \fstar and prove that it corresponds to the rules; i.e.,
that for all input configurations $C$, \ls{step}$(C) = C'$ implies
that $C \rightarrow C'$ according to the
semantics. Then, the core of each principal's implementation is
an \fstar{} stub function \kw{tstep} that repeatedly invokes \ls{step} on the
AST of the source program (produced by the \fstar extractor run
in a custom mode), unless the AST is an \kw{as_sec}
node. Functions \ls{step} and \kw{tstep} are extracted to OCaml by the
standard \fstar{} extraction process.

Local evaluation is not defined for \ls$as_sec$, so the stub
implements what amounts to {\sc{P-enter}} and {\sc{P-exit}} from
Figure~\ref{fig:dsl-proto-semantics}. When the stub notices the
program has reached an \kw{as_sec} expression, it calls into a circuit
library we have written that converts the AST of the second argument
of \kw{as_sec} to a boolean circuit. This circuit and the encoded inputs are
communicated to a co-hosted server that implements the GMW \mc
protocol~\cite{cryptoeprint:2011:257}. 
The server evaluates the circuit, coordinating with the GMW
servers of the other principals, and sends back the result. The
circuit library decodes the result and returns it to the stub.
The stub then carries on with the local evaluation. Our FFI interface
currently provides a form of monomorphic, first-order
interoperability between the (dynamically typed) interpreter and the
host language.

Our \fstar{} formalization of the \ourlang semantics, including the AST
specification, is 1900 lines of code. This formalization is used both
by the metatheory as well as by the (executable) interpreter. The
metatheory that connects the ST and DS semantics
(\S\ref{sec:formal}) is 3000 lines. The interpreter
and its correctness proof are another 290 lines of \fstar{} code.
The interpreter \kw{step} function is essentially a
big switch-case on the current expression, that calls into the
functions from the semantics specification. The \kw{tstep} stub is
another 15 lines. The size of the circuit
library, not including the GMW implementation, is 836 lines.
The stub, the implementation of GMW, the circuit library, and \fstar
toolchain (including the custom \ourlang extraction mode) are part
of our Trusted Computing Base (TCB).

\section{Applications}
\label{sec:apps}

In addition to joint median, presented in \S\ref{sec:overview}, we have
implemented and proved properties of two other \mc applications,
\emph{dealing for on-line card games} and \emph{private set intersection} (PSI).

\Paragraph{Card dealing} We have implemented an \mc-based card
dealing application in \ourlang. Such an application can play the
role of the dealer in a game of online poker, thereby eliminating the
need to trust the game portal for card dealing. The application relies
on \ourlang's support for \emph{secret
shares}~\citep{Shamir79}. Using secret shares, the participating
parties can share a value in a way that none of the parties can
observe the actual value individually (each party's share consists of
some random-looking bytes), but they can recover the value by combining
their shares in \ls{sec} mode.

In the application, the parties maintain a list of secret shares of
already dealt cards (the number of already dealt cards is public
information). To deal a new card, each party first generates a random
number locally. The parties then perform a secure computation to
compute the sum of their random numbers modulo 52, let's call it
$n$. The output of the secure computation is secret shares of $n$. Before
declaring $n$ as the newly dealt card, the parties needs to ensure
that the card $n$ has not already been dealt. To do so, they
iterate over the list of secret shares of already dealt cards, and for
each element of the list, check that it is different from $n$. The
check is performed in a secure computation that simply combines the shares
of $n$, combines the shares of the list element, and checks the
equality of the two values. If $n$ is different from all the
previously dealt cards, it is declared to be the new card, else the
parties repeat the protocol by again generating a fresh random number
each.

\ourlang provides the following API for secret shares:

\begin{lstlisting}
type Sh: Type -> Type
type can_sh: Type -> Type
assume Cansh_int: can_sh int

val v_of_sh: sh:Sh $\alpha$ -> Ghost $\alpha$
val ps_of_sh: sh:Sh $\alpha$ -> Ghost prins

val mk_sh: x:$\alpha$ -> Wys (Sh $\alpha$)
    (requires (fun m -> m.mode = Sec /\ can_sh $\alpha$))
    (ensures (fun m r tr -> v_of_sh r = x /\ ps_of_sh r = m.ps /\ tr = [])
val comb_sh: x:Sh $\alpha$ -> Wys $\alpha$ (requires (fun m -> m.mode = Sec /\ ps_of_sh x = m.ps))
                             (ensures (fun m r tr -> v_of_sh x = r /\ tr = [])
\end{lstlisting}

Type \ls{Sh $\alpha$} types the shares of values of type \ls{$\alpha$}. Our
implementation currently supports shares of \ls{int} values only;
the \ls{can_sh} predicate enforces this restriction on the source
programs. Extending secret shares support to other
types (such as pairs) should be straightforward (as in~\cite{wysteria}).
Functions \ls{v_of_sh}
and \ls{ps_of_sh} are marked \ls{Ghost}, meaning that they can only
be used in specifications for reasoning purposes. In the concrete
code, shares are created and combined using the \ls{mk_sh}
and \ls{comb_sh} functions. Together, the specifications of these
functions enforce that the shares are created and combined by the same
set of parties (through \ls{ps_of_sh}), and that \ls{comb_sh} recovers
the original value (through \ls{v_of_sh}).
The \ourlang interpreter transparently handles the low-level details of
extracting shares from the GMW implementation of Choi et
al. (\ls{mk_sh}), and reconstituting the shares back (\ls{comb_sh}).

In addition to implementing the card dealing application
in \ourlang, we have formally verified that the
returned card is fresh. The signature of the function that checks for
freshness of the newly dealt card is as follows (\ls{abc} is the set
of three parties in the computation):

\begin{lstlisting}
val check_fresh: l:list (Sh int){forall s'. mem s' l ==> ps_of_sh s' = abc}
  -> s:Sh int{ps_of_sh s = abc}
  -> Wys bool (requires (fun m -> m = Mode Par abc))
    (ensures (fun _ r _ -> r <==> (forall s'. mem s' l ==> not (v_of_sh s' = v_of_sh s))))
\end{lstlisting}

The specification says that the function takes two arguments: \ls{l}
is the list of secret shares of already dealt cards, and \ls{s} is the
secret shares of the newly dealt card. The function returns a boolean
\ls{r} that is \ls{true} iff the concrete value (\ls{v_of_sh})
of \ls{s} is different from the concrete values of all the elements
of the list \ls{l}. Using \fstar{}, we verify that the implementation
of \ls{check_fresh} meets this specification.

\Paragraph{PSI}
Consider a dating application that enables its users to compute their
common interests without revealing all of them.
This is an instance of the more general private set intersection
(PSI) problem~\cite{Huang12}.

We implement a straightforward version of PSI in \ourlang:

\begin{lstlisting}[frame=single]
let psi a b (input_a:sealed {a} (list int)) (input_b:sealed {b} (list int)) (l_a:int) (l_b:int) = 
  as_sec {a,b} (fun () -> List.intersect (reveal input_a) (reveal input_b) l_a l_b)
\end{lstlisting}

\noindent where the input sets are expressed as lists with public lengths.

Huang et al.~\cite{Huang12} provide an optimized PSI algorithm
that performs much better when the density of common elements in
the two sets is high. We implement their algorithm in \ourlang.
The optimized version consists of two nested loops -- an outer loop for Alice's
set and an inner loop for Bob's -- where an iteration of the inner loop
compares the current element of Alice's set with the current element of
Bob's. The nested loops are written using \ls{as_par} so that both Alice and
Bob execute the loops in lockstep (note that the set sizes are public), while
the comparison in the inner loop happens using \ls{as_sec}.
Instead of naive \ls{l_a * l_b} comparisons, Huang et al.~\cite{Huang12} observe
that once an element of Alice's set \ls{ax}
matches an element of Bob's set \ls{bx}, the inner loop can return immediately,
skipping the comparisons of \ls{ax} with the rest of Bob's set. Furthermore,
\ls{bx} can be removed from Bob's set, excluding it from any further comparisons
with other elements in Alice's set. Since there are no repeats in the input sets,
all the excluded comparisons are guaranteed to be false. We show the full code
and its performance comparison with \ls{psi} in
Appendix A.

As with the median example from \S\ref{sec:overview}, the optimized PSI
intentionally reveals more for performance gains. As such, we would like
to verify that the optimizations do not reveal more about parties' inputs.
We take the following stepwise refinement approach. First, we characterize the trace of
the optimized implementation as a pure function \ls{trace_psi_opt la lb}
(omitted for space reasons),
and show that the trace of \ls{psi_opt} is precisely \ls{trace_psi_opt la lb}.

Then, we define an intermediate PSI implementation that has the same
nested loop structure, but performs
\ls{l_a * l_b} comparisons without any optimizations. We characterize
the trace of this intermediate implementation as the pure function
\ls{trace_psi}, and show that it precisely captures the trace.

To show that \ls{trace_psi} does not reveal more than the intersection of
the input sets, we prove the following lemma. 

\newcommand\defeq{\mathrel{\overset{\makebox[0pt]{\mbox{\normalfont\tiny\sffamily def}}}{=}}}

\begin{lstlisting}
$\Psi$ la$_0$ la$_1$ lb$_0$ lb$_1$ $\defeq$ (* possibly diff input sets, but with *)
  la$_0$ $\cap$ lb$_0$ = la$_1$ $\cap$ lb$_1$ /\ (* intersections the same *)
  length la$_0$ = length la$_1$ /\ length lb$_0$ = length lb$_1$ (* lengths the same *)

val psi__interim_is_secure: la$_0$:_ -> lb$_0$:_ -> la$_1$:_ -> lb$_1$:_ -> Lemma
  (requires ($\Psi$ la$_0$ la$_1$ lb$_0$ lb$_1$)) (ensures (permutation (trace_psi la$_0$ lb$_0$) (trace_psi la$_1$ lb$_1$)))
\end{lstlisting}

The lemma essentially says that for two runs on same length inputs,
if the output is the same, then the resulting traces are permutation of
each other.\footnote{Holding Bob's
  (resp. Alice's) inputs fixed and varying Alice's (resp. Bob's) inputs,
  as done for \ls$median$ in \S\ref{sec:verification},
  is covered by this more general property.}
We can reason about the traces of \ls$psi_interim$ up to
permutation because Alice has no prior knowledge of the choice of
representation of Bob's set (Bob can shuffle his list), so cannot
learn anything from a permutation of the trace.\footnote{We could formalize
this observation using a probabilistic, relational variant
of \fstar~\citep{rfstar14}.} This establishes the security of
\ls{psi_interim}.

Finally, we can connect \ls$psi_interim$ to \ls$psi_opt$
by showing that there exists a function \ls$f$, such that for
any trace \ls$tr=trace_psi la lb$, the trace of \ls$psi_opt$,
\ls$trace_psi_opt la lb$, can be computed by
\ls$f (length la) (length lb) tr$.
In other words, the trace produced by the optimized implementation can be computed
using a function of information already available to Alice (or Bob)
when she (or he) observes a run of the secure, unoptimized version
\ls$psi_interim la lb$. As such, the optimizations do not reveal
further information.



\section{Related work}
\label{sec:related}

\Paragraph{Source \mc verification} While the verification of the
underlying crypto protocols has received some attention
\cite{Almeida:2017:FVS:3133956.3134017,cryptoeprint:2014:456},
the verification of correctness and security properties of \mc source
programs has remained largely unexplored, surprisingly so given that
the goal of \mc is to preserve the privacy of secret inputs.
The only previous work that
we know of is Backes et. al. \cite{backes_et_al:LIPIcs:2010:2877} who
devise an applied pi-calculus based abstraction for \mc, and use it
for formal verification. For an auction protocol that computes the
\kw{min} function, their abstraction comprises about 1400 lines of
code. \ourlang, on the other hand, enables direct verification of the
higher-level \mc source programs, and not their models,
and in addition provides a partially verified toolchain.

\Paragraph{Wysteria}
\ourlang's computational model is based on programming
abstractions of a previous domain-specific language,
Wysteria~\citep{wysteria}. \ourlang's realization as an embedded DSL in
\fstar{} makes important advances. In particular, \ourlang (a) enhances the
Wysteria semantics to include a notion of observable traces, and provides
the novel capability to prove security and correctness properties
about mixed-mode \mc source programs, (b) expands the
programming constructs available by drawing on features and libraries
of \fstar, and (c) adds assurance via a (partially) proved-correct
implementation.

\Paragraph{Verified \mc toolchain} Almeida et al.~\cite{Almeida:2017:FVS:3133956.3134017}
build a verified toolchain consisting of (a) a verified
circuit compiler from (a subset of) C to boolean circuits, and
(b) a verified implementation of Yao's~\cite{Yao}
garbled circuits protocol for 2-party \mc.
They use CompCert~\cite{Leroy2009Formal}
for the former, and EasyCrypt~\cite{Barthe:2011:CSP:2033036.2033043}
for the latter. These are significant advances, but there are several distinctions
from our work.
The MPC programs in their toolchain are not \emph{mixed-mode},
and thus it cannot express examples like \ls{median_opt} and the optimized PSI.
Their framework does not enable formal verification of source programs
like \ourlang does. It may be possible to use other frameworks for
verifying C programs (e.g. Frama-C~\cite{framac}), but it is inconvenient
as one has to work in the subset of C that falls in the intersection
of these tools. \ourlang is also more general as it supports
general $n$-party MPC; e.g., the card dealing application in \S\ref{sec:apps}
has 3 parties. Nevertheless, \ourlang may use the
verified Yao implementation for the special case of 2 parties.

\Paragraph{\mc DSLs and DSL extensions} In addition to Wysteria
several other \mc DSLs have been proposed in the
literature~\cite{Huang11,viff,Malka2011,fairplaymp,Holzer12,Nielsen07,Nielsen09,sharemind,Schropfer2011,wysteria,Liu2014,Laud:2015:DLL:2810103.2813664}. Most
of these languages have standalone implementations, and the
(usability/scalability) drawbacks
that come with them. Like \ourlang,
a few are implemented as language extensions. Launchbury et
al.~\citet{Launchbury2012} describe a Haskell-embedded DSL for writing
low-level ``share protocols'' on a multi-server ``SMC machine''.
OblivC~\cite{oblivc} is an extension to C for two-party \mc that
annotates variables and conditionals with an \kw{obliv} qualifier to
identify private inputs; these programs are compiled by
source-to-source translation.
The former is essentially a shallow embedding, and the latter is
compiler-based; \ourlang is unique in that it combines a shallow embedding
to support source program
verification and a deep embedding to support a non-standard
target semantics. Recent work~\cite{ezpc,Buscher:2018:HCH:3243734.3243786} compiles
to cryptographic protocols that include both arithmetic and boolean
circuits; the compiler decides which fragments
of the program fall into which category. It would be interesting work
to integrate such a backend in \ourlang.



\Paragraph{Mechanized metatheory} Our
verification results are different from a typical verification result
that might either mechanize metatheory for
an idealized language~\cite{Aydemir:2005:MMM:2145056.2145062}, or
might prove an interpreter or compiler correct w.r.t. a formal
semantics~\cite{Leroy2009Formal}---we do both. We mechanize the 
metatheory of \ourlang establishing the soundness of the conceptual
ST semantics w.r.t. the actual DS semantics, and
mechanize the proof that the interpreter implements
the correct DS semantics. 

\Paragraph{General DSL implementation strategies} DSLs (for \mc or
other purposes) are implemented
in various ways, such as by developing a standalone
compiler/interpreter, or by shallow or deep embedding in
a host language. Our approach bears relation
to the approach taken in LINQ~\cite{Meijer:2006:LRO:1142473.1142552},
which embeds a query language in normal C\# programs, and implements
these programs by extracting the query syntax tree and passing it to a
\emph{provider} to implement for a particular backend. Other
researchers have embedded DSLs in verification-oriented host languages
(e.g., Bedrock~\cite{bedrock} in Coq~\cite{coq}) to permit formal
proofs of DSL programs. Low$^\star$~\cite{Protzenko:2017:VLP:3136534.3110261}
is a shallow-embedding of a small, sequential,
well-behaved subset of C in \fstar that extracts to C using a \fstar-to-C
compiler. Low$^\star$ has been used to verify and implement several cryptographic
constructions. Fromherz et al.~\cite{valepopl} present a deep embedding
of a subset of x64 assembly in \fstar that allows efficient verification of
assembly and its interoperation with C code generated from Low$^\star$.
They design (and verify) a custom VC generator for the deeply embedded DSL,
that allows for the proofs of assembly crypto routines to scale.

\section{Conclusions}

This paper has presented \ourlang, the first DSL to enable formal
verification of efficient source \mc programs as written in a
full-featured host programming language, \fstar. The paper presented
examples such as joint median, card dealing, and PSI, and showed
how the DSL enables their correctness and security proofs.
\ourlang implementation, examples, and
proofs are publicly available on
\iflong
Github at \url{https://github.com/FStarLang/FStar/tree/stratified\_last/examples/wysteria}.
\else
Github.
\fi

\bibliographystyle{splncs04}
\bibliography{secure-computation,../fstar}


\section{Appendix A: Optimized PSI}
\label{sec:appendixa}

\begin{lstlisting}[xleftmargin=1em, numbers=left, frame=single]
let rec for_each_alice a b la lb = (* outer loop to iterate over Alice's list la *)
  if la=[] then []
  else let lb, r = check_each_bob a b (List.hd la) lb in
      r::for_each_alice a b (List.tl la) lb
and check_each_bob a b ax lb  = (* for Alice's element ax, check its matches in lb *)
  if lb=[] then [], []
  else let bx = List.hd lb in
      let r = as_sec {a,b} (fun () -> reveal ax = reveal bx) in $\label{line:gmw-v0}$
      if r then List.tl lb, [r]  $\label{line:v1-opt}$ (* optimization: skip rest of lb, and remove bx from lb *)
      else let lb', r' = check_each_bob a b ax (List.tl lb) in
          bx::lb', r::r'
let psi_opt a b (la:list (sealed {a} int)) (lb:list (sealed {b} int)) = as_par {a,b} ( $\label{line:psiopt}$
  fun () -> let bs = build_matrix (for_each_alice a b la lb) in $\label{line:bs-v0}$
           let ia = as_par {a} (fun () -> filteri (contains true $\circ$ row bs) la) in$\label{line:ia}$
           let ib = as_par {b} (fun () -> filteri (contains true $\circ$ col bs) lb) in$\label{line:ib}$
           concat (mkmap a ia) (mkmap b ib)) $\label{line:give}$
\end{lstlisting}

The function \ls$psi_opt$ (line~\ref{line:psiopt})
begins by calling \ls$for_each_alice$, in \ls{par} mode, which in turn calls
\ls$check_each_bob$, for each element \ls$ax$ of Alice's list \ls$la$.
The function \ls$check_each_bob$ (which is also run by both Alice and
Bob, as they are in \ls{par} mode) iterates over Bob's list \ls{lb}, and for each element
\ls{bx}, it performs a secure computation to compare \ls{ax} and \ls{bx}
(line~\ref{line:gmw-v0}).
Importantly, at line~\ref{line:v1-opt}, once we
detect that \ls$ax$ is in the intersection, we return
immediately instead of comparing \ls$ax$ against the remaining
elements of \ls$lb$. Furthermore, we remove \ls$bx$ from \ls$lb$,
excluding it from any future comparisons with other elements of
Alice's set \ls$la$. Since \ls$la$ and \ls$lb$ are representations of
sets (no repeats), all the excluded comparisons are
guaranteed to be false. Once all the comparisons are accumulated
(as a matrix), Alice and Bob inspect them
(line~\ref{line:ia} and line~\ref{line:ib}) to determine
which of their elements are in the
intersection.

\begin{figure}
\begin{small}
  \includegraphics[scale=0.68]{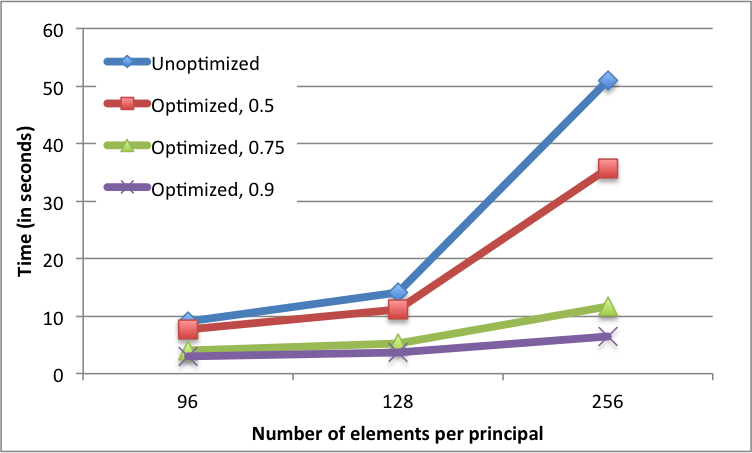}
  \caption{Time to run (in secs) normal and optimized PSI for varying per-party set sizes and intersection densities.}
  \label{fig:psibench}
  \end{small}
\end{figure}

We evaluate the performance of
the \kw{psi} (computing intersection in a single secure
computation), and the \kw{psi_opt} (the optimized version) algorithms
from \S\ref{sec:apps}. The programs that we benchmark are
slightly different than the ones presented there, in that the
local \kw{col} and \kw{row} functions are not the verified ones. The
results are shown in Figure~\ref{fig:psibench}. We measure the time (in
seconds) for per party set sizes 96, 128, and 256, and intersection
densities (i.e. the fraction of elements that are common) 0.5, 0.75,
and 0.9.

The time taken by the unoptimized version is independent of the
intersection density since it always compares all pairs of
values. However, as the intersection density increases, the optimized
version performs far better -- it is able to skip many
comparisons. For lower densities ($<$ 0.35), the optimization does not
improve performance, as the algorithm essentially becomes quadratic,
and the setup cost for each secure computation takes over.

We note that a similar performance profile was also noted by Rastogi
et. al. \cite{wysteria}, although they did not experiment with set
size 256, like we did.

\section{Appendix B: Formalization}

\begin{figure*}[t]
\[
\begin{array}{l}
\inferrule*[lab=S-aspar]
{
e_1 = \kw{as_par}\;\ps\;(L_1, \lambda x.e) \quad M = \kw{Par}\;\ps_1 \quad \ps \subseteq \ps_1 \\\\
X_1 = (M; L; \kw{seal}\;\ps\;\langle\rangle; T), X
}
{
M; X; L; T; e_1 \rightarrow \kw{Par}\;\ps;X_1; L_1[x \mapsto ()]; \cdot; e
}
\hspace{0.3cm}
\inferrule*[lab=S-parret]
{
X = (M_1; L_1; \kw{seal}\;\ps\;\langle\rangle; T_1), X_1\\\\
\kw{can_seal}\;\ps\;v \quad
T_2 = \kw{append}\;T_1\;[\kw{TScope}\;\ps\;T]
}
{
M; X; L; T; v \rightarrow M_1; X_1; L_1; T_2; \kw{sealed}\;\ps\;v
}
\\\\
\inferrule*[lab=S-assec]
{
e_1 = \kw{as_sec}\;\ps\;(L_1, \lambda x.e) \quad M = \kw{Par}\;\ps \\\\
X_1 = (M; L; \langle\rangle\; T), X
}
{
M; X; L; T; e_1  \rightarrow \kw{Sec}\;\ps; X_1; L_1[x \mapsto ()]; \cdot; e
}
\hspace{0.9cm}
\inferrule*[lab=S-secret]
{
\kw{is_sec}\;M \quad X = (M_1; L_1; \langle\rangle; T), X_1 \\\\
T_1 = \kw{append}\;T\;[\kw{TMsg}\;v] \quad
}
{
M; X; L; \cdot; v \rightarrow M_1; X_1; L_1; T_1; v
}
\\\\
\inferrule*[lab=S-seal]
{
M = \_\;\ps_1 \quad \ps \subseteq \ps_1
}
{
M; X; L; T; \kw{seal}\;\ps\;v \rightarrow M; X; L; T; \kw{sealed}\;\ps\;v
}
\hspace{0.2cm}
\inferrule*[lab=S-reveal]
{
M = \kw{Par}\;\ps_1 \Rightarrow \ps_1 \subseteq \ps \quad\quad
M = \kw{Sec}\;\ps_1 \Rightarrow \ps_1 \cap \ps \neq \phi
}
{
M; X; L; T; \kw{reveal}\;(\kw{sealed}\;\ps\;v) \rightarrow M; X; L; T; v
}
\\\\
\inferrule*[lab=S-mkmap]
{
M = \kw{Par}\;\ps_1 \Rightarrow v = \kw{sealed}\;\ps_2\;v_2 \wedge \ps \subseteq \ps_1 \wedge \ps \subseteq \ps_2\\\\
M = \kw{Sec}\;\ps_1 \Rightarrow \ps \subseteq \ps_1 \wedge v_2 = v
}
{
M; X; L; T; \kw{mkmap}\;\ps\;v \rightarrow M; X; L; T; [\ps \mapsto v_2]
}
\hspace{0.5cm}
\inferrule*[lab=S-proj]
{
M = \kw{Par}\;\ps \Rightarrow \ps = \kw{singleton}\;p \\\\
M = \kw{Sec}\;\ps \Rightarrow p \in \ps \quad
m[p] = v
}
{
M; X; L; T; \kw{project}\;m\;p \rightarrow M; X; L; T; v
}
\\\\
\inferrule*[lab=S-concat]
{
\mathsf{dom}(m_1) \cap \mathsf{dom}(m_1) = \phi
}
{
M; X; L; T; \kw{concat}\;m_1\;m_2 \rightarrow M; X; L; T; m_1 \uplus m_2
}
\hspace{1.5cm}
\inferrule*[lab=S-ffi]
{
v = \kw{exec_ffi}\;\ext{f}\;\bar{v}
}
{
M; X; L; T; \kw{ffi}\;\ext{f}\;\bar{v} \rightarrow M; X; L; T; v
}
\\\\
\end{array}
\]
\caption{\ourlang ST semantics (selected rules)}
\label{fig:src-semantics-appen}
\end{figure*}

\begin{figure*}[t]
\[
\begin{array}{l}
\inferrule*[lab=L-let]
{
X_1 = (M; L; \kw{let}\;x\;=\;\langle\rangle\;\kw{in}\;e_2; T), X
}
{
M; X; L; T; \kw{let}\;x\;=\;e_1\;\kw{in}\;e_2 \rightarrow M; X_1; L; \cdot; e_1
}
\hspace{0.2cm}
\inferrule*[lab=L-app]
{
}
{
M; X; L; T; (L_1, \lambda x.e)\;e_1 \rightarrow M; X; L_1[x \mapsto e_1]; T; e
}
\\\\
\inferrule*[lab=L-aspar1]
{
e_1 = \kw{as_par}\;\ps\;(L_1, \lambda x.e) \quad p \in \ps \\\\
X_1 = (M; L; \kw{seal}\;\ps\;\langle\rangle; T), X
}
{
\kw{Par}\;p; X; L; T; e_1 \leadsto \kw{Par}\;p; X_1; L_1[x \mapsto ()]; \cdot; e
}
\hspace{0.3cm}
\inferrule*[lab=L-parret]
{
X = (M; L_1; \kw{seal}\;\ps\;\langle\rangle; T_1), X_1 \\\\
T_2 = \kw{append}\;T_1\;T \quad v_1 = \kw{sealed}\;\ps\;v
}
{
\kw{Par}\;p; X; L; T; v \leadsto \kw{Par}\;p; X_1; L_1; T_2; v_1
}
\\\\
\hspace{1cm}\inferrule*[left=L-aspar2]
{
p \not\in \ps
}
{
\kw{Par}\;p; X; L; T; \kw{as_par}\;\ps\;(L_1, \lambda x.e) \leadsto \kw{Par}\;p; X; L; T; \kw{sealed}\;\ps\;\bullet
}
\\\\
\inferrule*[lab=L-seal]
{
p \in \ps \Rightarrow v_1 = \kw{seal}\;\ps\;v \quad
p \not\in \ps \Rightarrow v_1 = \kw{seal}\;\ps\;\bullet
}
{
M; X; L; T; \kw{seal}\;\ps\;v \leadsto M; X; L; T; v_1
}
\hspace{0.2cm}
\inferrule*[lab=L-reveal]
{
p \in \ps
}
{
M; X; L; T; \kw{reveal}\;(\kw{sealed}\;\ps\;v) \leadsto M; X; L; T; v
}
\\\\
\inferrule*[lab=L-mkmap]
{
v = \kw{sealed}\;\ps_2\;v_2 \\\\
p \in \ps \Rightarrow p \in \ps_2 \wedge m = [p \mapsto v] \quad p \not\in \ps \Rightarrow m = \cdot
}
{
M; X; L; T; \kw{mkmap}\;\ps\;v \rightarrow M; X; L; T; m
}
\hspace{0.2cm}
\inferrule*[lab=L-proj]
{
p = p_1 \quad m = [p \mapsto v]
}
{
M; X; L; T; \kw{project}\;m\;p_1 \rightarrow M; X; L; T; v
}
\\\\
\hspace{1.8cm}
\inferrule*[left=L-concat]
{
\mathsf{dom}(m_1) \cap \mathsf{dom}(m_1) = \phi
}
{
M; X; L; T; \kw{concat}\;m_1\;m_2 \rightarrow M; X; L; T; m_1 \uplus m_2
}
\end{array}
\]
\caption{Distributed semantics, selected local rules {\small{(the mode $M$ is always \kw{Par}\;$p$)}}}
\label{fig:dsl-tgt-semantics-appen}
\end{figure*}

\begin{figure*}
\footnotesize

\begin{lstlisting}
type as_mode = | Par | Sec
type mode = | Mode: m:as_mode -> ps:prins -> mode
type telt =
  | TMsg  : #a:Type -> x:a -> telt
  | TScope: ps:prins -> t:list telt -> telt

type trace = list telt
effect Wys (a:Type) (req:mode -> Type) (ens:mode -> a -> trace -> Type)

val as_sec: ps:prins -> f:(unit -> Wys a pre post) -> Wys a (requires (fun m -> m=Mode Par ps /\ pre (Mode Sec ps))) (ensures  (fun m r tr -> tr=[TMsg r] /\ post (Mode Sec ps) r [])))

val as_par: ps:prins -> f:(unit -> Wys a pre post) -> Wys (sealed ps a) (requires (fun m -> m.mode=Par /\ ps $\subseteq$ m.ps /\ can_seal ps a /\ pre (Mode Par ps))) (ensures  (fun m r tr -> exists t. tr=[TScope ps t] /\ post (Mode Par ps) (unseal r) t)))

type sealed : prins -> Type -> Type
val unseal: #ps:prins -> sealed ps $\alpha$ -> Ghost $\alpha$

val seal: ps:prins -> x:$\alpha$ -> Wys (sealed ps $\alpha$) (requires (fun m -> ps $\subseteq$ m.ps)) (ensures (fun m r tr -> x=unseal r /\ tr=[]))

val reveal: #ps:prins -> x:sealed ps $\alpha$ -> Wys $\alpha$ (requires (fun m -> m.mode=Par ==> m.ps $\subseteq$ ps /\ m.mode=Sec ==> m.ps $\cap$ ps $\neq$ $\emptyset$)) (ensures (fun m r tr -> r=unseal a /\ tr=[]))

type map : prins -> Type -> Type
val mkmap_p: #ps$_1$:prins -> eps:eprins -> x:sealed $\alpha$ ps$_1$ -> Wys (map $\alpha$ eps) (requires (fun m -> m.mode=Par /\ eps $\subseteq$ ps$_1$ /\ eps $\subseteq$ m.ps)) (ensures (fun m r tr -> r = const_map eps (unseal x) /\ tr = []))

val mkmap_s: eps:eprins -> x:$\alpha$ -> Wys (map $\alpha$ eps) (requires (fun m -> m.mode=Sec /\ eps $\subseteq$ m.ps)) (ensures (fun m r tr -> r = const_map eps x /\ tr = []))

val project: #eps:eprins -> p:prin -> x:map $\alpha$ eps{contains p x} -> Wys $\alpha$ (requires (fun m -> m.mode=Par ==> m.ps = singleton p /\ m.mode=Sec ==> mem p m.ps)) (ensures (fun m r tr -> r = select p x /\ tr = []))

val concat: #eps$_x$:eprins -> #eps$_y$:eprins -> x:map $\alpha$ eps$_x$ -> y:map $\alpha$ eps$_y$ -> Wys (map $\alpha$ (eps$_x$ $\cup$ eps$_y$)) (requires (fun m -> disjoint (dom x) (dom y))) (ensures (fun m r tr -> r = join x y /\ tr = [])

type Sh: Type -> Type
type can_sh: Type -> Type
assume Cansh_int: can_sh int

val v_of_sh: #a:Type -> sh:Sh a -> Ghost a
val ps_of_sh: #a:Type -> sh:Sh a -> Ghost prins
val mk_sh: #a:Type -> x:a -> Wys (Sh a) (requires (fun m -> m.mode = Sec /\ can_sh a)) (ensures (fun m r tr -> v_of_sh r = x /\ ps_of_sh r = m.ps /\ tr = [])

val comb_sh: #a:Type -> x:Sh a -> Wys a (requires (fun m -> m.mode = Sec /\ ps_of_sh x = m.ps)) (ensures (fun m r tr -> v_of_sh x = r /\ tr = [])
\end{lstlisting}
\caption{\ourlang API}
\label{fig:wysstar-api}
\end{figure*}

\end{document}